\documentstyle[epsfig,a4]{article}

\begin{document}

\renewcommand{\thesection}{\arabic{section}}
\renewcommand{\figurename}{\small Fig.}
\renewcommand{\theequation}{\arabic{section}.\arabic{equation}}

\newcommand{\eqreset}{\setcounter{equation}{0}}

\newcommand{\gtrsim}{
\,\raisebox{0.35ex}{$>$}
\hspace{-1.7ex}\raisebox{-0.65ex}{$\sim$}\,
}

\newcommand{\lesssim}{
\,\raisebox{0.35ex}{$<$}
\hspace{-1.7ex}\raisebox{-0.65ex}{$\sim$}\,
}

\newcommand{\onehalf}{\mbox{\scriptsize 
\raisebox{1.5mm}{1}\hspace{-2.7mm}
\raisebox{0.5mm}{$-$}\hspace{-2.8mm}
\raisebox{-0.9mm}{2}\hspace{-0.7mm}
\normalsize }}

\newcommand{\onefourth}{\mbox{\scriptsize 
\raisebox{1.5mm}{1}\hspace{-2.7mm}
\raisebox{0.5mm}{$-$}\hspace{-2.8mm}
\raisebox{-0.9mm}{4}\hspace{-0.7mm}
\normalsize }}

\bibliographystyle{prsty}

\begin{flushleft}

{\small 
 J. Phys.: Condens. Matter (to appear)}
\vspace{2cm}

{
\Large\bf
Field dependence of the temperature at the peak of the ZFC magnetization
}      

\vspace{0.1cm}

\end{flushleft}

\begin{flushright}
\parbox[t]{11cm}{
H\ Kachkachi$^{1,*}$, W\ T \ Coffey$^{2}$,
D\ S\ F\ Crothers$^{3}$, A\ Ezzir$^{1}$, E\ C\ Kennedy$^{3}$, M\ Nogu\`{e}s$%
^{1},$ and E\ Tronc$^{4}$
\vspace{0.1cm}

\small

$^{1}$ Lab. de Magn\'{e}tisme et d'Optique, CNRS, Univ. de Versailles St. Quentin, \\
45, av. des Etats-Unis, 78035 Versailles CEDEX, France\\
$^{2}$ Dept. of Electronics \& Electrical Engineering, Trinity College,\\
Dublin 2, Ireland\\
$^{3}$Dept. of Applied Mathematics \& Theoretical Physics, The Queen's Univ.%
\\
of Belfast, Belfast BT7 1NN, Northern Ireland\\
$^{4}$ LCMC, CNRS, Univ. Pierre\ et Marie\ Curie, 75252 Paris Cedex 05,\\
France
\vspace{1cm}

Version as of \today

\vspace{0.4cm}

{\bf Abstract.} \hspace{1mm}
The effect of an applied magnetic field on the temperature at the maximum of
the ZFC magnetization, $M_{ZFC},$ is studied using the recently obtained
analytic results of Coffey et al. (Phys.\ Rev.\ Lett., {\bf 80 }(1998) 5655)
for the prefactor of the N\'{e}el relaxation time which allow one to
precisely calculate the prefactor in the N\'{e}el-Brown model and thus the
blocking temperature as a function of the coefficients of the Taylor series
expansion of the magnetocrystalline anisotropy. The present calculations
indicate that even a precise determination of the prefactor in the
N\'{e}el-Brown theory, which always predicts a monotonic decrease of the
relaxation time with increasing field, is insufficient to explain the effect
of an applied magnetic field on the temperature at the maximum of the ZFC
magnetization. On the other hand, we find that the non linear
field-dependence of the magnetization along with the magnetocrystalline
anisotropy appears to be of crucial importance to the existence of this
maximum.
\vspace{2cm}
} 

\end{flushright}

\renewcommand{\thefootnote}{\fnsymbol{footnote}}
\renewcommand{\footnoterule}{\rule{0cm}{0cm}}
\footnotetext[1]{Corresponding author: kachkach@physique.uvsq.fr}

\section{Introduction}
\begin{figure}[t]
\unitlength1cm
\begin{picture}(15,12)
\centerline{\epsfig{file=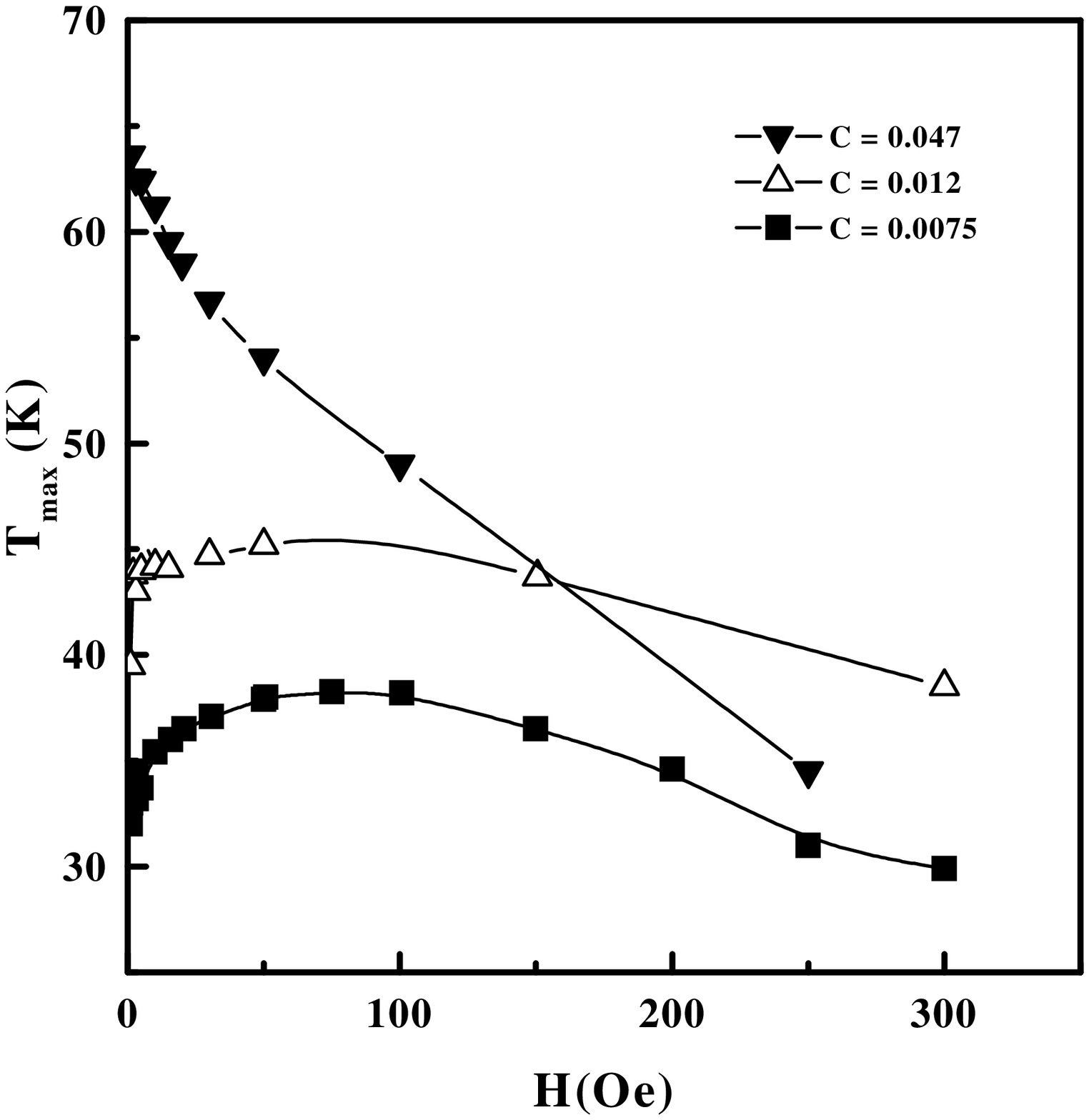,angle=0,width=10cm}}
\end{picture}
\vspace{-5cm}
\caption{ \label{fig1}
The temperature $T_{\max }$ at the maximum of the ZFC
magnetization plotted against the applied field for different sample
concentrations. The volume fraction of the particles in the sample 4D
($\gamma-Fe_{2}O_{3}$, mean diameter $\sim 8.32$ nm)   
determined from the density measurements is $C_{v}=0.0075$, $0.012$, $0.047$.}
\end{figure}

\eqreset
Experimental results obtained a few years ago for ferrofluids \cite{Litt}
and recently for $\gamma $-Fe$_{2}$O$_{3}$ nanoparticles \cite{Sappey}
indicate that for dilute samples (weak interparticle interactions), the
temperature $T_{\max }$ at the maximum of the zero-field-cooled
magnetization $M_{ZFC},$ first increases with increasing field, attains a
maximum and then decreases. More recently, additional experiments performed
on the $\gamma $-Fe$_{2}$O$_{3}$ particles dispersed in a polymer \cite
{Ezzir} confirm the previous result for dilute samples and show that, on
the contrary, for concentrated samples (strong interparticle interactions) $%
T_{\max }$ is a monotonic decreasing function of the magnetic field \cite
{Ezzir} (see Fig. \ref{fig1}).
The behaviour observed for dilute samples (isolated nanoparticles) is
rather unusual since one intuitively expects the applied field to lower the
energy barrier and thus the blocking temperature of all particles and
thereby the temperature $T_{\max }$ \footnote{
The temperature $T_{\max }$ at the maximum of the ZFC magnetization is
assumed to be roughly given by the average of the blocking temperature $%
T_{B} $ of all particles in the (dilute) assembly.}. Resonant tunneling is
one of the arguments which has been advanced \cite{Friedman} as a mechanism
responsible for an increase of the relaxation rate around zero field. More
recently $M_{ZFC}$ measurements \cite{Sappey} have shown an anomaly in the
temperature range 40-60K, which is probably too high for quantum effects to
manifest themselves. Yet another explanation \cite{Sappey} of the $T_{\max }$
effect was proposed using arguments based on the particle anisotropy-barrier
distribution. It was suggested that for randomly oriented particles of a 
{\it uniform size}, and for small values of the field, the field depresses
the energy barriers, and thereby enlarges the barrier distribution, so
lowering the blocking temperature. It was also suggested that the increase
of the barrier-distribution width overcompensates for the decrease of the
energy barrier in a single particle. However, the discussion of the
relaxation time was based on the original Arrhenius approach of N\'{e}el.
Here the escape rate, that is the inverse relaxation time, is modelled as an
attempt frequency multiplied by the Boltzmann factor of the barrier height,
with the inverse of the attempt frequency being of the order of 10$^{-10}$
s, thus precluding any discussion of the effect of the applied field and the
damping parameter on the prefactor, and so their influence on the blocking
temperature.

It is the purpose of this paper to calculate the blocking temperature using
the Kramers theory of the escape of particles over potential barriers as
adapted to magnetic relaxation by Brown \cite{Brown}. The advantage of using
this theory is that it allows one to precisely calculate the prefactor as a
function of the applied field and the damping parameter (provided that
interactions between particles are neglected). Thus the behaviour of the
blocking temperature as a function of these parameters may be determined. It
appears from the results that even such a precise calculation is unable to
explain the maximum in the blocking temperature $T_{B}$ as a function of the
applied field. Thus an explanation of this effect is not possible using the
N\'{e}el-Brown model for a {\it single} particle as that model invariably
predicts a monotonic decrease of $T_{B}$ as a function of the applied field.

In view of this null result we demonstrate that the $T_{\max }$ effect may
be explained by considering an assembly of non-interacting particles having
a volume distribution. This is accomplished by using Gittleman's \cite
{Gittleman} model which consists of writing the zero-field cooled
magnetization of the assembly as a sum of two contributions, one from the
blocked magnetic moments and the other from the superparamagnetic ones, with
the crucial assumption that the superparamagnetic contribution is given by a
non-linear function (Langevin function) of the applied magnetic field and
temperature. If this is done even the simple N\'{e}el-Brown expression for
the relaxation time leads to a maximum in $T_{\max }$ for a wide range of
values of the anisotropy constant $K.$ It was claimed in \cite{Luc Thomas}
that a simple volume distribution, together with a N\'{e}el expression for
the relaxation time, leads to the same result for FeC particles, although
the author used the Langevin function for the superparamagnetic contribution
to magnetization.

Therefore, the particular expression for the single-particle relaxation time
which is used appears not to be of a crucial importance in the context of
the calculation of the blocking temperature.

In the next section we briefly review Kramers' theory of the escape rate.
\section{Kramers' escape rate theory}

The simple Arrhenius calculation of reaction rates for an assembly of {\it %
mechanical particles} undergoing translational Brownian motion, in the
presence of a potential barrier, was much improved upon by Kramers \cite
{Kramers}. He showed, by using the theory of the Brownian motion, how the
prefactor of the reaction rate, as a function of the damping parameter and
the shape of the potential well, could be calculated from the underlying
probability-density diffusion equation in phase space, which for Brownian
motion is the Fokker-Planck equation (FPE). He obtained (by linearizing the
FPE about the potential barrier) explicit results for the escape rate for
intermediate-to-high values of the damping parameter and also for very small
values of that parameter. Subsequently, a number of authors (\cite
{Meshkov-Melnikov}, \cite{BHL}) showed how this approach could be extended
to give formulae for the reaction rate which are valid for all values of the
damping parameter. These calculations have been reviewed by H\"{a}nggi et al.%
\cite{Hanggi et al.}.

The above reaction-rate calculations pertain to an assembly of mechanical
particles of mass $m$ moving along the $x$-axis so that the Hamiltonian of a
typical particle is 
\begin{equation}
H=\frac{p^{2}}{2m}+V(q),  \label{Ham}
\end{equation}
where $q$ and $p$ are the position and momentum of a particle; and $V(q)$ is
the potential in which the assembly of particles resides, the interaction of
an individual particle with its surroundings is then modelled by the
Langevin equation 
\begin{equation}
\dot{p}+\varsigma p+\frac{dV}{dq}=\lambda (t)  \label{Langevin}
\end{equation}
where $\lambda (t)$ is the Gaussian white noise, and $\varsigma $ is the
viscous drag coefficient arising from the surroundings of the particle.

The Kramers theory was first adapted to the thermal rotational motion of the
magnetic moment (where the Hamiltonian, unlike that of Eq.(\ref{Ham}), is
effectively the Gibbs free energy) by Brown \cite{Brown} in order to improve
upon N\'{e}el's concept of the superparamagnetic relaxation process (which
implicitly assumes discrete orientations of the magnetic moment and which
does not take into account the gyromagnetic nature of the system). Brown in
his first explicit calculation \cite{Brown} of the escape rate confined
himself to axially symmetric (functions of the latitude only) potentials of
the magneto-crystalline anisotropy so that the calculation of the relaxation
rate is governed (for potential-barrier height significantly greater than $%
k_{B}T$) by the smallest non-vanishing eigenvalue of a one-dimensional
Fokker-Planck equation. Thus the rate obtained is valid for all values of
the damping parameter. As a consequence of this very particular result, the
analogy with the Kramers theory for mechanical particles only becomes fully
apparent when an attempt is made to treat non axially symmetric potentials
of the magneto-crystalline anisotropy which are functions of both the
latitude and the longitude. In this context, Brown \cite{Brown} succeeded in
giving a formula for the escape rate for magnetic moments of single-domain
particles, in the intermediate-to-high (IHD) damping limit, which is the
analogue of the Kramers IHD formula for mechanical particles. In his second
1979 calculation \cite{Brown} Brown only considered this case. Later Klik
and Gunther \cite{Klik and Gunther}, by using the theory of first-passage
times, obtained a formula for the escape rate which is the analogue of the
Kramers result for very low damping. All these (asymptotic) formulae which
apply to a general non-axially symmetric potential, were calculated
explicitly for the case of a uniform magnetic field applied at an arbitrary
angle to the anisotropy axis of the particle by Geoghegan et al. \cite
{Geoghegan et al} and compare favorably with the exact reaction rate given
by the smallest non-vanishing eigenvalue of the FPE \cite{Coffey1}%
,\thinspace \cite{Coffey2}, \cite{Coffey3} and with experiments on the
relaxation time of single-domain particles \cite{Wensdorfer et al.}.

In accordance with the objective stated in the introduction, we shall now
use these formulae (as specialized to a uniform field applied at an
arbitrary angle by Geoghegan et al. \cite{Geoghegan et al} and Coffey et al. 
\cite{Coffey1},\thinspace \cite{Coffey2}, \cite{Coffey3}) for the
calculation of the blocking temperature of a single particle.

A valuable result following from these calculations will be an explicit
demonstration of the breakdown of the non-axially symmetric asymptotic
formulae at very small departures from axial symmetry, manifesting itself in
the form of a spurious increase in $T_{\max }$. Here interpolation formulae
joining the axially symmetric and non-axially symmetric asymptotes
(analogous to the one that joins the oscillatory and non-oscillatory
solutions of the Schr\"{o}dinger equation in the WKBJ method \cite{Fermi})
must be used in order to reproduce the behaviour of the exact reaction rate
given by the smallest non-vanishing eigenvalue of the FPE, which always
predicts a monotonic decrease of $T_{\max }$, as has been demonstrated by
Garanin et al. \cite{Garanin et al.} in the case of a transverse field.

\section{Calculation of the blocking temperature}
\label{Calculation of the blocking temperature}

Following the work of Coffey et al. cited above, the effect of the applied
magnetic field on the blocking temperature is studied by extracting $T_{B}$
from the analytic (asymptotic) expressions for the relaxation-time (inverse
of the Kramers escape rate) \cite{Coffey1},\thinspace \cite{Coffey2}, \cite
{Coffey3}, which allow one to evaluate the prefactor as a function of the
applied field and the dimensionless damping parameter $\eta _{r}$ in the
Gilbert-Landau-Lifshitz (GLL) equation. For single-domain particles the
equation of motion of the unit vector describing the magnetization inside
the particle is regarded as the Langevin equation of the system (detailed in 
\cite{Coffey4}).

Our discussion of the N\'{e}el-Brown model as applied to the problem at hand
proceeds as follows~:

In an assembly of ferromagnetic particles with uniaxial anisotropy excluding
dipole-dipole interactions, the ratio of the potential energy $vU$ to the
thermal energy $k_{B}T$ of a particle is described by the bistable form 
\begin{equation}
\beta U=-\alpha ({\bf e\cdot n)}^{2}-\xi ({\bf e\cdot h)}  \label{1}
\end{equation}
where $\beta =v/(k_{B}T)$, $v$ is the volume of the single-domain particle; $%
\alpha =\beta K\gg 1$ is the anisotropy (reduced) barrier height parameter; $%
K$ is the anisotropy constant; $\xi =\beta M_{s}H$ is the external field
parameter; ${\bf e,n,}$ and ${\bf h}$ ($h\equiv \xi /2\alpha $) are unit
vectors in the direction of the magnetization ${\bf M}$, the easy axis, and
the magnetic field ${\bf H}$, respectively. $\theta $ and $\psi $ denote the
angles between ${\bf n}$ and ${\bf e}$ and between ${\bf n}$ and ${\bf h,}$
respectively. The N\'{e}el time, which is the time required for the magnetic
moment to surmount the potential barrier given by (\ref{1}), is
asymptotically related to the smallest nonvanishing eigenvalue $\lambda _{1}$
(the Kramers' escape rate) of the Fokker-Planck equation, by means of the
expression $\tau \approx 2\tau _{N}/_{\lambda _{1}}\cite{Brown}$, where the
diffusion time is
\begin{equation}
\tau _{N}\simeq \frac{\beta M_{s}}{2\gamma }\left[ \frac{1}{\eta _{r}}+\eta
_{r}\right] ,  \label{2}
\end{equation}
$\gamma $ is the gyromagnetic ratio, $M_{s}$ the intrinsic magnetization, $%
\eta $ the phenomenological damping constant, and $\eta _{r}$ the GLL
damping parameter $\eta _{r}=\eta \gamma M_{s}.$

As indicated above, Brown \cite{Brown} at first derived a formula for $%
\lambda _{1}$, for an arbitrary {\it axially symmetric} bistable potential
having minima at $\theta =(0,\pi )$ separated by a maximum at $\theta _{m},$
which when applied to Eq.(\ref{1}) for ${\bf h\Vert n,}$ i.e. a magnetic
field parallel to the easy axis, leads to the form given by Aharoni \cite
{Aharoni}, $\theta _{m}=\cos ^{-1}(-h),$%
\begin{equation}
\lambda _{1}\approx \frac{2}{\sqrt{\pi }}\alpha ^{3/2}(1-h^{2})\times \left[
(1+h)\;e^{-\alpha (1+h)^{2}}+(1-h)\;e^{-\alpha (1-h)^{2}}\right]  \label{4}
\end{equation}
where $0\leq h\leq 1,$ $h=1$ being the critical value at which the bistable
nature of the potential disappears.

In order to describe the non-axially symmetric asymptotic behaviour, let us
denote by $\beta \Delta U_{-}$ the smaller reduced barrier height of the two
constituting escape from the left or the right of a bistable potential. Then
for very low damping, i.e. for $\eta _{r}\times \beta \Delta U_{-}\ll 1$
(with of course the reduced barrier height $\beta \Delta U_{-}\gg 1$,
depending on the size of the nanoparticle studied) we have \cite{Coffey3}, 
\cite{Brown}, \cite{Coffey4} the following asymptotic expression for the
N\'{e}el relaxation time 
\begin{eqnarray}
\tau _{VLD}^{-1} &\approx &\frac{\lambda }{2\tau _{N}}  \label{LD} \\
&\approx &\frac{\eta _{r}}{2\pi }\left\{ \omega _{1}\times \beta
(U_{0}-U_{1})e^{-\beta (U_{0}-U_{1})}+\omega _{2}\times \beta
(U_{0}-U_{2})e^{-\beta (U_{0}-U_{2})}\right\}  \nonumber
\end{eqnarray}
For the intermediate-to-high damping, where $\eta _{r}\times \beta \Delta
U_{-}>1$ (again with the reduced barrier height $\beta \Delta U_{-}$ much
greater than unity) we have \cite{Coffey3} the asymptotic expression 
\begin{equation}
\tau _{IHD}^{-1}\approx \frac{\Omega _{0}}{2\pi \omega _{0}}\left\{ \omega
_{1}e^{-\beta (U_{0}-U_{1})}+\omega _{2}e^{-\beta (U_{0}-U_{2})}\right\} ,
\label{IHD}
\end{equation}
where 
\begin{eqnarray*}
\omega _{1}^{2} &=&\frac{\gamma ^{2}}{M_{s}^{2}}c_{1}^{(1)}c_{2}^{(1)},\quad
\omega _{2}^{2}=\frac{\gamma ^{2}}{M_{s}^{2}}c_{1}^{(2)}c_{2}^{(2)} \\
\omega _{0}^{2} &=&-\frac{\gamma ^{2}}{M_{s}^{2}}c_{1}^{(0)}c_{2}^{(0)},\quad
\\
\Omega _{0} &=&\frac{\eta _{r}g^{\prime }}{2}\left[ -c_{1}^{(0)}-c_{2}^{(0)}+%
\sqrt{(c_{2}^{(0)}-c_{1}^{(0)})^{2}-\frac{4}{\eta _{r}^{2}}%
c_{1}^{(0)}c_{2}^{(0)}}\right] \\
g^{\prime } &=&\frac{\gamma }{(1+\eta _{r}^{2})M_{s}}
\end{eqnarray*}
Here $\omega _{1},\omega _{2}$ and $\omega _{0}$ are respectively the well
and saddle angular frequencies associated with the bistable potential, $%
\Omega _{0}$ is the damped saddle angular frequency and the $c_{j}^{(i)}$
are the coefficients of the truncated (at the second order in the direction
cosines) Taylor series expansion of the crystalline anisotropy and external
field potential at the wells of the bistable potential denoted by $1$ and $2$
and at the saddle point denoted by $0$. A full discussion of the application
of these general formulae to the particular potential, which involves the
numerical solution of a quartic equation in order to determine the $%
c_{j}^{(i)}$ with the exception of the particular field angle $\psi =\frac{%
\pi }{4}$ or $\frac{\pi }{2}$, in Eq.(\ref{1}) is given in Refs. \cite
{Geoghegan et al}, \cite{Kennedy}.

The blocking temperature $T_{B}$ is defined as the temperature at which $%
\tau =\tau _{m}$, $\tau _{m}$ being the measuring time. Therefore, using
Eqs.(\ref{2}), (\ref{4}) and (\ref{LD}) (or (\ref{2}), (\ref{4}) and (\ref
{IHD})) and solving the equation $\tau =\tau _{m}$ for the blocking
temperature $T_{B}$, we obtain the variation of $T_{B}$ as a function of the
applied field, for an arbitrary angle $\psi $ between the easy axis and the
applied magnetic field.

\begin{figure}[t]
\unitlength1cm
\begin{picture}(15,11)
\centerline{\epsfig{file=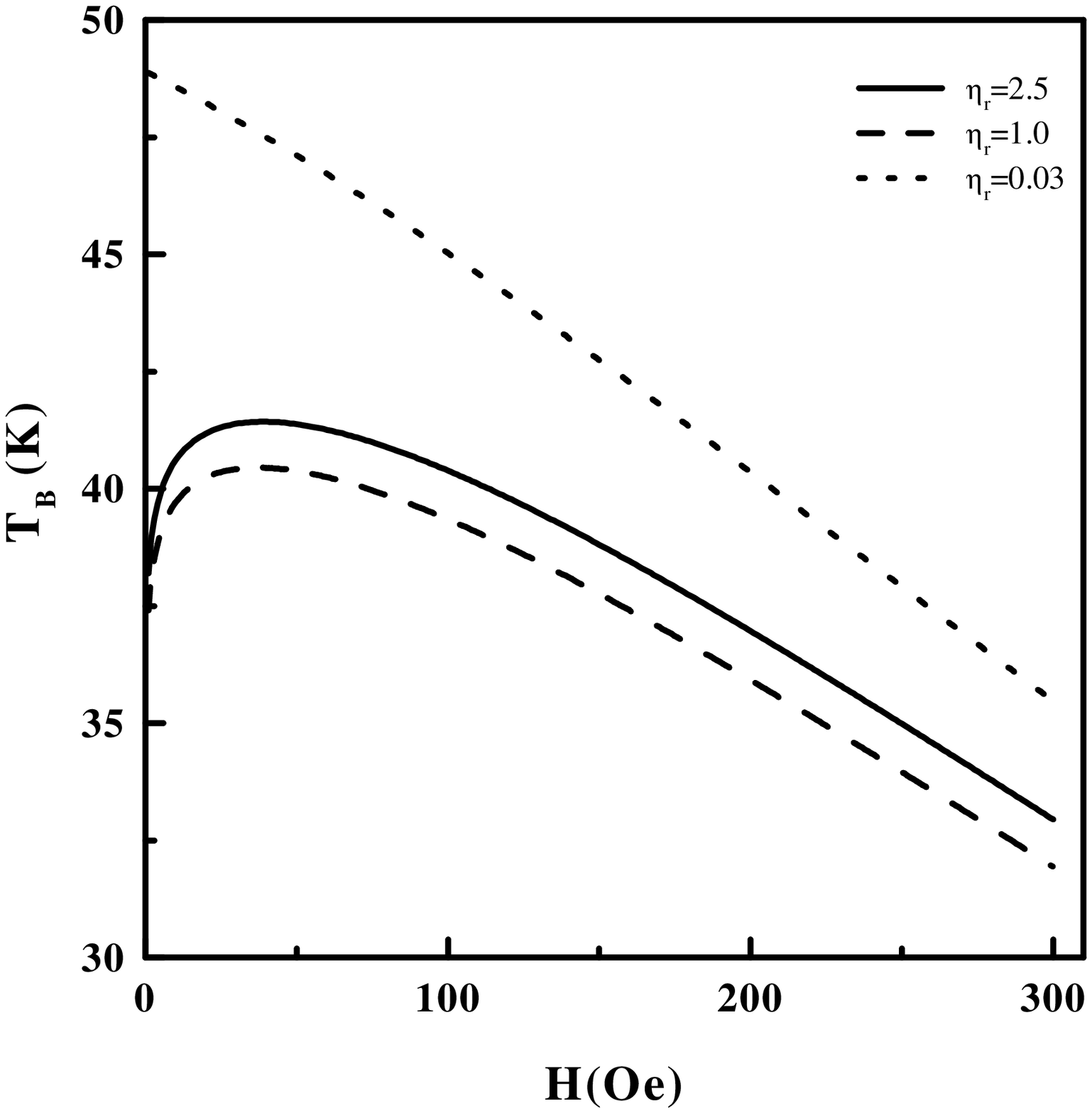,angle=0,width=10cm}}
\end{picture}
\vspace{-4cm}
\caption{ \label{fig2}
The blocking temperature $T_{B}$ as a function of the
applied field as extracted from the formulae for the relaxation time of a
single-domain particle and summed over the arbitrary angle $\psi ,$ plotted
for different values of the damping parameter $\eta _{r}$ (see text), and
mean volume $\left\langle V\right\rangle =$ $265$ nm$^{3}.$ ($K=1.25\times
10^{5}$erg/cm$^{3},\;\gamma =2.10^{7}$ S$^{-1}$G$^{-1},M_{s}=300$ emu/cm$%
^{3},\;\tau _{m}=100$ s).
}
\end{figure}

\begin{figure}[t]
\unitlength1cm
\begin{picture}(15,11)
\centerline{\epsfig{file=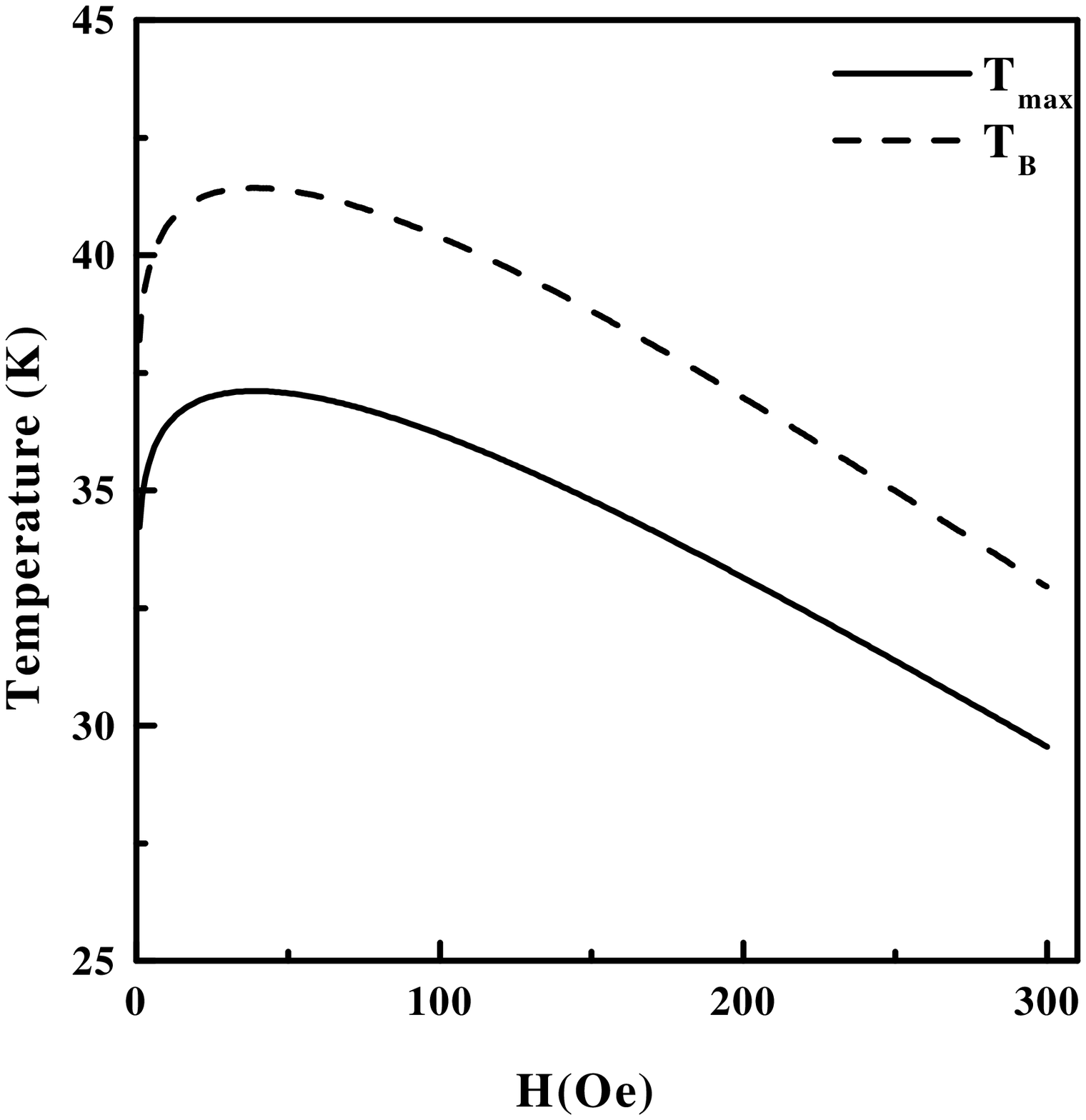,angle=0,width=10cm}}
\end{picture}
\vspace{-4cm}
\caption{ \label{fig3}
The temperature $T_{\max }$ at the maximum of the ZFC
magnetization for an assembly of particles with randomly distributed easy
axes, as a function of the applied field obtained by averaging the blocking
temperature in Fig.\ 2 over the volume log-normal distribution ($\sigma
=0.85 $), see text; with $\eta _{r}\medskip =2.5$.
}
\end{figure}

In particular, for very small values of $\psi $ we have used Eq.(\ref{4}),
as the problem then becomes almost axially symmetric and the arguments
leading to Eqs.(\ref{LD}) and (\ref{IHD}) fail \cite{Brown}, \cite{Geoghegan
et al}, \cite{Coffey4}, and appropriate connection formulae must be used so
that they may attain the axially symmetric limit. We then sum over $\psi ,$
as the easy axes of the particles in the assembly are assumed to be randomly
distributed. In Fig. \ref{fig2} we have plotted the resulting $T_{B}$ vs. $H$ for
different values of the damping parameter $\eta _{r}$. We have checked that
lowering (or raising) the value of the measuring time $\tau _{m}$ shifts the
curve $T_{B}(H)$ only very slightly upwards (or downwards) while leaving the
qualitative behaviour unaltered. 
In order to compare our analytical results
with those of experiments on particle assemblies, we have calculated the
temperature $T_{\max }$ at the maximum of the ZFC magnetization. In the
present calculations we have assumed that $M_{s}$ is independent of
temperature. We find that the temperature $T_{\max }$ behaves in the same
way as was observed experimentally \cite{Sappey}, \cite{Ezzir} for dilute
samples (see Fig.\ \ref{fig3}, where the parameters are those of the most dilute sample
in Fig.\ \ref{fig1}, with $\eta _{r}=2.5$).
%
%
\noindent Moreover, our calculations for a single particle show that the
blocking temperature $T_{B}(H)$ exhibits a bell-like shape in the case of
intermediate-to-high damping. This behaviour is however spurious as is shown
below.

\section{Spurious behaviour of the blocking temperature at low fields}

We have mentioned that initially the non-axially symmetric asymptotic
formulae appear to render the low-field behaviour of $T_{\max }$. However,
this apparent behaviour is spurious as the asymptotic formulae (as one may
verify (i) by exact calculation of the smallest non-vanishing eigenvalue of
the Fokker-Planck equation, and (ii) e.g. the IHD formula does not continue
to the axially symmetric asymptote) fail at low fields, since the IHD
formula diverges like $h^{-1/2},$ for all angles, where $h$ is the reduced
field as defined in sect.\ \ref{Calculation of the blocking
temperature}. The effect of the divergence is thus to 
produce a spurious maximum in $T_{\max }$ as a function of the applied field.

In order to verify this, we have also performed such calculations \cite
{Geoghegan et al}, \cite{Coffey1}, \cite{Kennedy} using exact numerical
diagonalization of the Fokker-Planck matrix. The smallest non-vanishing
eigenvalue $\lambda _{1}$ thus obtained leads to a blocking temperature
which agrees with that rendered by the asymptotic formulae with the all
important exception of IHD at very low fields where the exact calculation
invariably predicts a monotonic decrease in the blocking temperature rather
than the peak predicted by the IHD formula (\ref{IHD}), so indicating that
the theoretical peak is an artefact of the asymptotic expansion, caused by
using Eq.(\ref{IHD}) in a region beyond its range of validity, that is in a
region where the potential is almost axially symmetric due to the smallness
of the applied field which gives rise to a spurious discontinuity between
the axially and non axially symmetric asymptotic formulae.

An explanation of this behaviour follows (see also \cite{Garanin et al.}):
in the non-axially symmetric IHD asymptote Eq.(\ref{IHD}) which is
formulated in terms of the Kramers escape rate, as the field tends to zero,
for high damping, the saddle angular frequency $\omega _{0}$ tends to zero.
Thus the saddle becomes infinitely wide and so the escape rate predicted by
Eq.(\ref{IHD}) diverges leading to an apparent rise in the blocking
temperature until the field reaches a value sufficiently high to allow the
exponential in the Arrhenius terms to take over. When this occurs the
blocking temperature decreases again in accordance with the expected
behaviour. This is the field range where one would expect the non-axially
asymptote to work well.

In reality, as demonstrated by the exact numerical calculations of the
smallest non vanishing eigenvalue of the Fokker-Planck matrix, the small
field behaviour is not as predicted by the asymptotic behaviour of Eq.(\ref
{IHD}) (it is rather given by the axially-symmetric asymptote) because the
saddle is limited in size to $\omega _{0}.$ Thus the true escape rate cannot
diverge, and the apparent discontinuity between the axially-symmetric and
non axially-symmetric results is spurious, leading to apparent rise in $%
T_{B} $. In reality, the prefactor in Eq.(\ref{IHD}) can never overcome the
exponential decrease embodied in the Arrhenius factor. Garanin \cite
{Garanin2} (see \cite{Garanin et al.}) has discovered bridging formulae
which provide continuity between the axially-symmetric Eq.(\ref{4}) and non
axially symmetric asymptotes leading to a monotonic decrease of the blocking
temperature with the field in accordance with the numerical calculations of
the lowest eigenvalue of the Fokker-Planck equation.

An illustration of this was given in Ref. \cite{Garanin et al.} for the
particular case of $\psi =\frac{\pi }{2},$ that is a transverse applied
field. If the escape rate is written in the form 
\[
\tau ^{-1}=\frac{\omega _{1}}{\pi }A\exp (-\beta \Delta U) 
\]
where $\omega _{1}$ is the attempt frequency and is given by 
\[
\omega _{1}=\frac{2K\gamma }{M_{s}}\sqrt{1-h^{2}}, 
\]
then the factor $A,$ as predicted by the IHD formula, behaves as $\eta _{r}/%
\sqrt{h}$ for $h\ll 1,\eta _{r}^{2},$ while for $h=0$, $A$ behaves as $2\pi
\eta _{r}\sqrt{\alpha /\pi },$ which is obviously discontinuous. So that a
suitable interpolation formula is required. Such a formula (analogous to
that used in the WKBJ method \cite{Fermi}) is obtained by multiplying the
factor $A$ of the axially symmetric result by $e^{-\xi }I_{0}(\xi ),$ where $%
I_{0}(\xi )$ is the modified Bessel function of the first kind, and $\xi
=2\alpha h$ (see \cite{Garanin et al.}).

This interpolation formula, as is obvious from the large and small $\xi $
limits, automatically removes the undesirable $1/\sqrt{h}$ divergence of the
IHD formula and establishes continuity between the axially symmetric and
non-axially symmetric asymptotes for $\psi =\pi /2$, as dictated by the
exact solution.

It is apparent from the discussion of this section that the N\'{e}el-Brown
model for a single particle is unable to explain the maximum in $T_{\max },$
as a careful calculation of the asymptotes demonstrates that they always
predict a monotonic decrease in the blocking temperature. However this
effect may be explained by considering an assembly of non-interacting
particles with a (log-normal) volume distribution and using
Gittleman's \cite
{Gittleman} model as shown below, where the superparamagnetic contribution
to magnetization is taken to be a non-linear function (Langevin function) of
the magnetic field.

\section{Possible explanation of the maximum in $T_{\max }$}
\label{Possible explanation of the maximum}

Our explanation of the low-field behaviour of $T_{\max }$ is based on
extracting $T_{\max }$ from the zero-field cooled magnetization curve
assuming a volume distribution of particles. According to Gittleman's model
the zero-field cooled magnetization of the assembly can be written as a sum
of two contributions, one from the blocked magnetic moments and the other
from the superparamagnetic ones. In addition, we write the superparamagnetic
contribution as a Langevin function of the applied magnetic field and
temperature.

Gittleman \cite{Gittleman} proposed a model in which the alternative
susceptibility of an assembly of non-interacting particles, with a volume
distribution and randomly distributed easy axes, can be written as 
\begin{equation}
\chi (T,\omega )=\frac{1}{Z}%
\displaystyle \int %
\limits_{0}^{\infty }dVVf(V)\chi _{V}(T,\omega ),  \label{Susc1}
\end{equation}
where $Z=$ $\int_{0}^{\infty }dVVf(V),$ $f(V)$ is the volume distribution
function, $\chi _{V}$ is the susceptibility of the volume under
consideration, and $dVVf(V)\chi _{V}$ is the contribution to the total
susceptibility corresponding to volumes in the range $V-V+dV.$ $\chi _{V}$
is then calculated by assuming a step function for the magnetic field, i.e. $%
H=0$ for $t<0$ and $H=H_{0}=const$ for $t>0.$ Then, the contribution to the
magnetization from particles of volume $V$ is given by 
\begin{equation}
M_{V}(t)=VH_{0}\left( \chi _{0}-(\chi _{0}-\chi _{1})e^{-t/\tau }\right) ,
\label{Susc2}
\end{equation}
where $\chi _{0}=M_{s}^{2}(T)V/3k_{B}T$ is the susceptibility at
thermodynamic equilibrium and $\chi _{1}=M_{s}^{2}(T)V/3E_{B}$ is the
initial susceptibility of particles in the blocked state (see \cite{Dormann
et al.} and many references therein). The Fourier transform of (\ref{Susc2})
leads to the complex susceptibility 
\begin{equation}
\chi =\frac{(\chi _{0}+i\omega \tau \chi _{1})}{1+i\omega \tau },
\label{Susc3}
\end{equation}
whose real part reads \cite{Gittleman} 
\begin{equation}
\chi ^{\prime }=\frac{\chi _{0}+\omega ^{2}\tau ^{2}\chi _{1}}{1+\omega
^{2}\tau ^{2}},  \label{Susc4}
\end{equation}
where $\tau _{m}$ is the measuring time, and $\omega $ is the angular
frequency ($=2\pi \nu $).

Starting from (\ref{Susc4}) the application of an alternating field yields:
a) $\chi ^{\prime }=\chi _{0}$ if $\omega \tau \ll 1.$ At high temperature
the magnetic moments orientate themselves on a great number of occasions
during the time of a measurement, and thus the susceptibility is the
superparamagnetic susceptibility $\chi _{0}.$ b) $\chi ^{\prime }=\chi _{1}$
if $\omega \tau \gg 1.$ At low temperature the energy supplied by the field
is insufficient to reverse the magnetic moments the time of a measurement.
Here the susceptibility is the static susceptibility $\chi _{1}.$ Between
these two extremes there exists a maximum at the temperature $T_{\max }.$
$\chi ^{\prime }$ can be calculated from (\ref{Susc4}) using the formula for
the relaxation time $\tau $ appropriate to the anisotropy symmetry, and
considering a particular volume $V,$ one can determine the temperature $%
T_{\max }.$

In an assembly of particles with a volume distribution, $\chi^{\prime }$
can be calculated for a (large) volume distribution by postulating that at a
given temperature and given measuring time, certain particles are in the
superparamagnetic state and that the others are in the blocked state. The
susceptibility is then given by the sum of two contributions 
\begin{equation}
\chi ^{\prime }(T,\nu )=%
\displaystyle \int %
\limits_{V_{c}}^{\infty }dVVf(V)\chi _{1}(T,\nu )+%
\displaystyle \int %
\limits_{0}^{V_{c}}dVVf(V)\chi _{0}(T,\nu ),  \label{Susc5}
\end{equation}
where $V_{c}=V_{c}(T,H,\nu )$ is the blocking volume defined as the volume
for which $\tau =\frac{1}{\nu} = \tau_{m}$. $\chi ^{\prime }$ shows a maximum at $%
T_{\max }$ near $<T_{B}>.$

If this is done even the simple N\'{e}el-Brown\footnote{%
This is the simplest non-trivial case since the relaxation time (and thereby
the critical volume) depends on the magnetic field.} expression for the
relaxation time leads to a maximum in $T_{\max }$ when the superparamagnetic
contribution to magnetization is a Langevin function of the magnetic field.
Thus the particular expression for the single-particle relaxation time used
appears not to be of a crucial importance in the context of the calculation
of the blocking temperature.

In Figs. \ref{fig4a}-\ref{fig4c} we plot the result of such calculations (see appendix) where the
parameters correspond to the samples of Fig. \ref{fig1}. In Fig. \ref{fig4a} we compare the
results from linear and non-linear (Langevin function) dependence of the
magnetization on the magnetic field. We see that indeed the non-linear
dependence on $H$ of the superparamagnetic contribution to magnetization
leads to a maximum in $T_{\max }$ while in the linear case the temperature $%
T_{\max }$ is a monotonic function of the field, for all values of $K$
(corresponding to our samples). This only shows that the volume distribution
by itself cannot account for the non-monotonic behaviour of the temperature $%
T_{\max }$, contrary to what was claimed in \cite{Luc Thomas}.
%
\begin{figure}[t]
\unitlength1cm
\begin{picture}(15,12)
\centerline{\epsfig{file=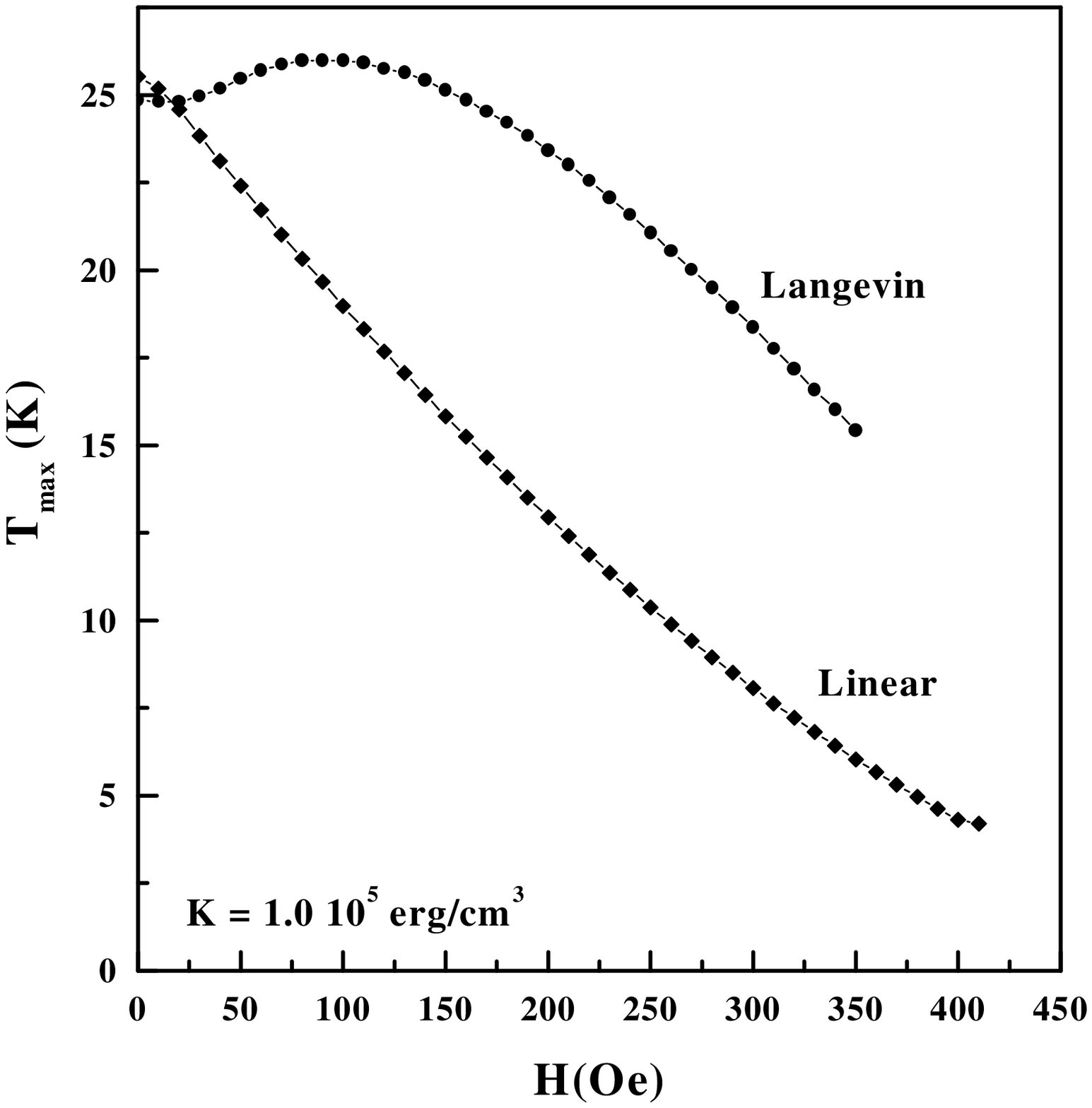,angle=0,width=10cm}}
\end{picture}
\vspace{-5cm}
\caption{ \label{fig4a}
Temperature $T_{\max }$ as a function of the applied field obtained by
the calculations of sect. V and appendix. Squares : the superparamagnetic
magnetization $M_{sp}$ is a linear function of the magnetic field given by
Eq.(\ref{11}). Circles : $M_{sp}$ is the Langevin function given by Eq.(\ref
{11b}). The parameters are the same as in Figs.1-3, and $K=1.0\times 10^{5}$%
erg/cm$^{3}$.
}
\end{figure}
%
%
\begin{figure}[t]
\unitlength1cm
\begin{picture}(15,11)
\centerline{\epsfig{file=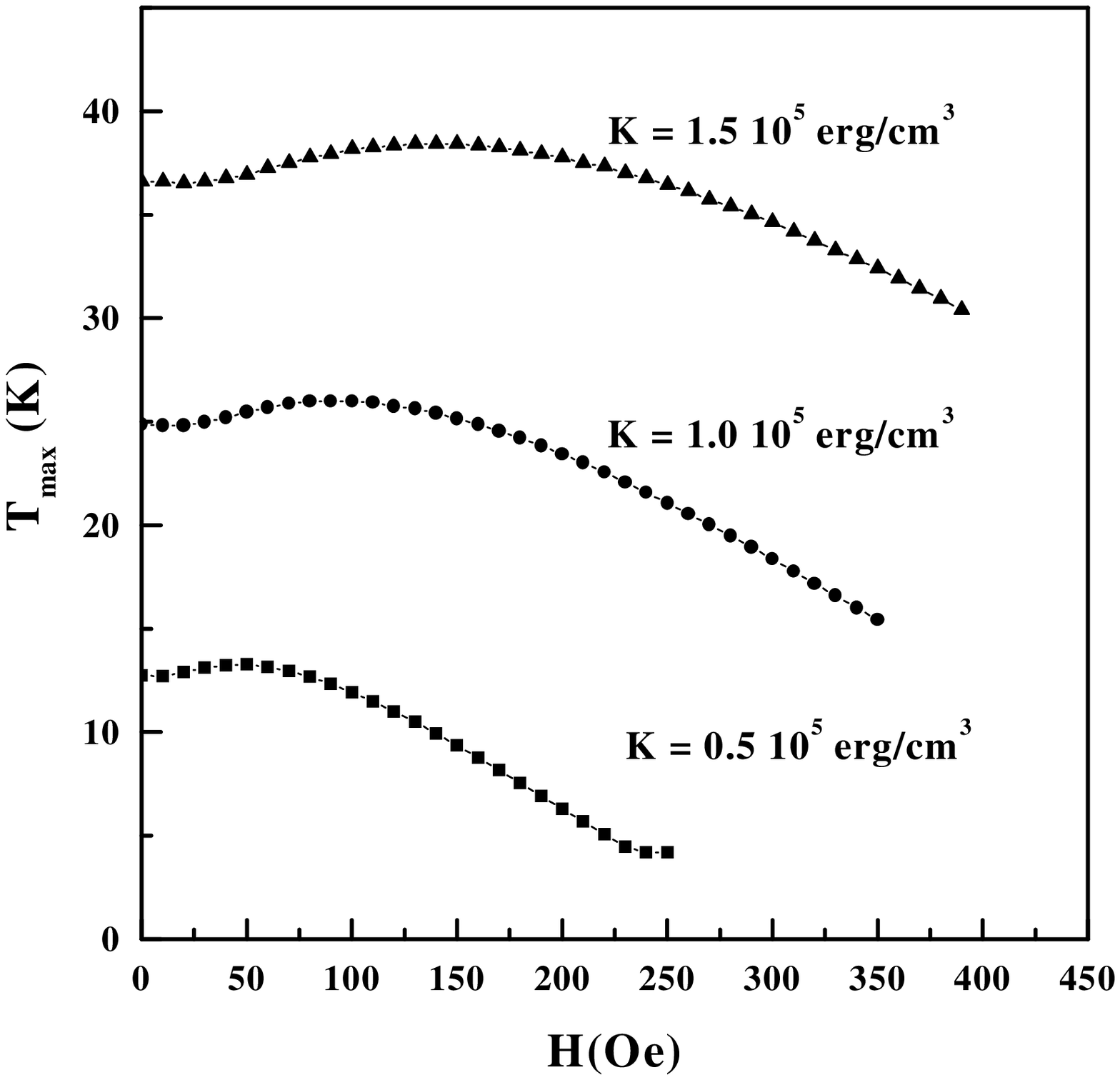,angle=0,width=9cm}}
\end{picture}
\vspace{-4.9cm}
\caption{ \label{fig4b}
Temperature $T_{\max }(H)$ for different values of the anisotropy
constant $K.$
}
\end{figure}
\begin{figure}[h]
\unitlength1cm
\begin{picture}(15,11.5)
\centerline{\epsfig{file=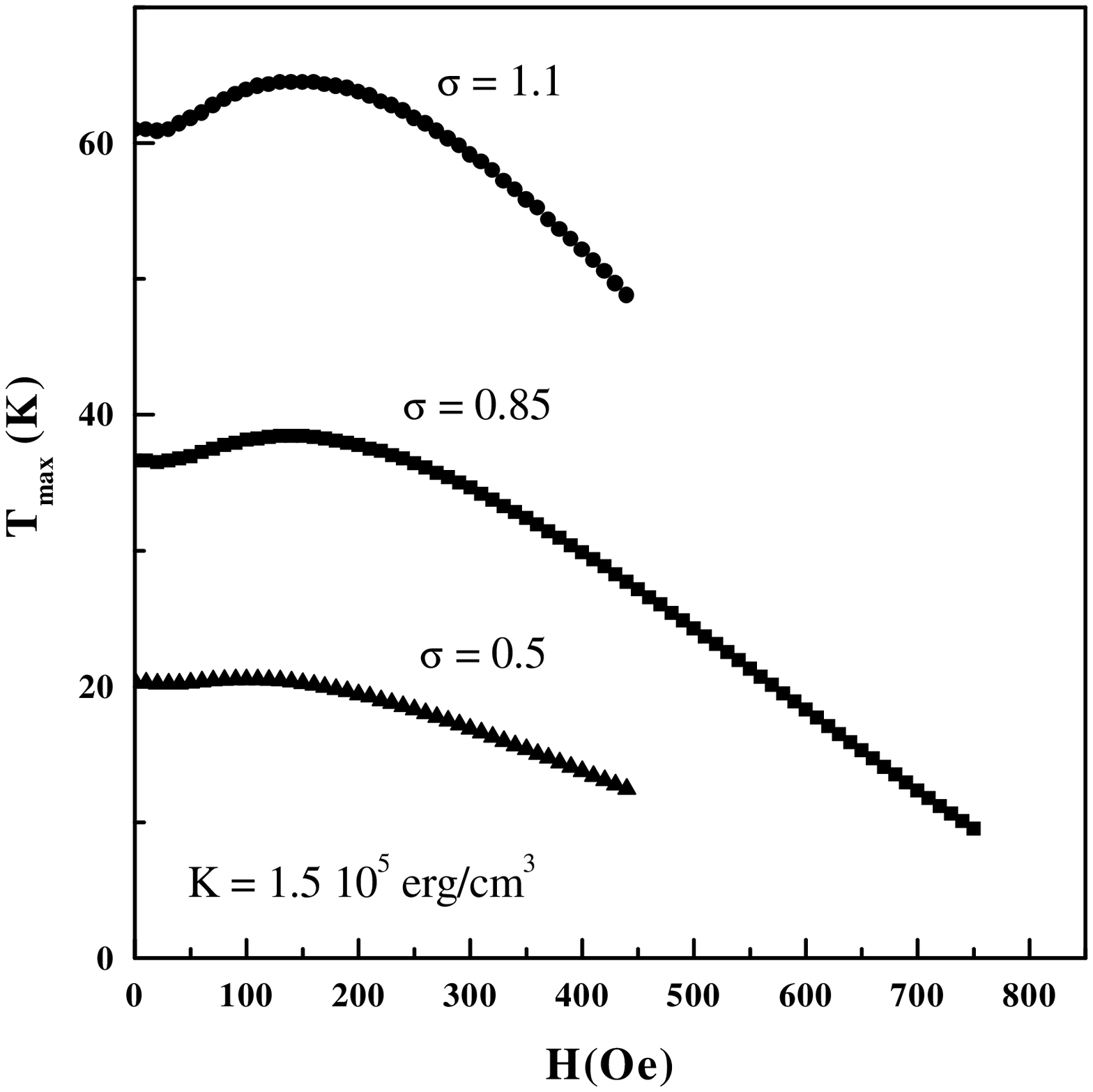,angle=0,width=8cm}}
\end{picture}
\vspace{-4cm}
\caption{ \label{fig4c}
$T_{\max }(H)$ for different values of the volume-distribution width $%
\sigma ,$ and $K=1.5\times 10^{5}$erg/cm$^{3}.$
}
\end{figure}

In fact, in the non-linear case, $T_{\max }$ exhibits three different
regimes with field ranges depending on all parameters and especially on $K$
(see Figs. \ref{fig4a}, \ref{fig4b}). For example, for $K=1.5\times 10^{5}$ erg/cm$^{3}$, the
field ranges are (in Oe) $0.0<H<30,\;30<H<140,$ and $H>140$. In the first
range, i.e. very low fields, $T_{\max }$ slightly decreases, then in the
second range it starts to increase up to a maximum, and finally for very
high fields $T_{\max }$ decreases down to zero. These three regimes were
obtained experimentally in \cite{Luc Thomas} in the case of diluted FeC
particles.

Next, we studied the effect of varying the anisotropy constant $K$. In
Fig. \ref{fig4b} we plot the temperature $T_{\max }$ vs. $H$ for different values of $%
K.$ It is seen that apart from the obvious shifting of the peak of $T_{\max
} $ to higher fields, this peak broadens as the anisotropy constant $K$
increases.

We have also varied the volume-distribution width $\sigma $ and the results
are shown in Fig. \ref{fig4c}. There we see that the maximum of $T_{\max }$ tends to
disappear as $\sigma $ becomes smaller.

Finally, these results show that the non-monotonic behaviour of $T_{\max }$
is mainly due to the non-linear dependence of the magnetization as a
function of magnetic field, and that the magneto-crystalline anisotropy and
the volume-distribution width have strong bearing on the variation of the
curvature of $T_{\max }$ vs. field.

\section{Conclusion}

Our attempt to explain the experimentally observed maximum in the curve $%
T_{\max }(H)$ for dilute samples using the asymptotic formulae for the
prefactor of the relaxation rate of a single-domain particle given by Coffey
et al. \cite{Coffey4}, has led to the conclusion that these asymptotic
formulae are not valid for small fields, where the maximum occurs. However,
this negative result has renewed interest in the long-standing problem of
finding bridging formulae between non-axially and axially symmetric
expressions for the prefactor of the escape rate. Recently, this problem has
been partially solved in \cite{Garanin et al.}.

On the other hand, exact numerical calculations \cite{Geoghegan et al}, \cite
{Coffey1}, \cite{Kennedy} of the smallest eigenvalue of the Fokker-Planck
matrix invariably lead to a monotonic decrease in the blocking temperature
(and thereby in the temperature $T_{\max }$) as a function of the magnetic
field. We may conclude then that the expression of the single-particle
relaxation time does not seem to play a crucial role. Indeed, the
calculations of sect.\ \ref{Possible explanation of the maximum} have
shown that even the simple N\'{e}el-Brown 
expression for the relaxation time leads to a maximum in $T_{\max }$ if one
considers an assembly of particles whose magnetization, formulated through
Gittleman's model, has a superparamagnetic contribution that is a Langevin
function of the magnetic field. The magneto-crystalline anisotropy and the
volume-distribution width have strong influence.

Another important point, whose study is beyond the scope of this work, is
the effect of interparticle interactions on the maximum in the temperature $%
T_{\max }$. As was said in the introduction, this maximum disappears in
concentrated samples, i.e. in the case of intermediate-to-strong
interparticle interactions. A recent study \cite{Chantrell} based on Monte
Carlo simulations of interacting (cobalt) fine particles seems to recover
this result but does not provide a clear physical interpretation of the
effect obtained. In particular, it was shown there that interactions have a
strong bearing on the effective variation of the average energy barrier with
field, as represented in an increase of the curvature of the variation of $%
T_{\max }$ with $H$ as the packing density (i.e. interparticle interactions)
increases.

\clearpage

\section*{Acknowledgements}
H.K., W.T.C. and E.K. thank D.\ Garanin for helpful conversations.
This work was supported in part by Forbairt Basic Research Grant
No SC/97/70 and the Forbairt research collaboration fund 1997-1999. The work
of D.S.F.C. and E.C.K. was supported by EPSRC (GR/L06225).
\newpage 

\section*{Appendix: Obtaining $T_{\max }$ from the ZFC magnetization}

Here we present the (numerical) method of computing the temperature $T_{\max
}$ at the peak of the zero-field cooled magnetization of non-interacting
nanoparticles.

The potential energy for a particle reads 
\begin{equation}
\frac{\beta U}{\alpha }=\sin ^{2}\theta -2h\cos (\psi -\theta )  \label{1app}
\end{equation}
where all parameters are defined in sect.\ \ref{Calculation of the blocking temperature}.

Then, one determines the extrema of the potential $U$ and defines the escape
rate $\lambda $ according to the symmetry of the problem. Here we consider,
for simplicity, the axially-symmetry N\'{e}el-Brown model where $\lambda $
is given by Eq. (\ref{4}).

The next step consists in finding the critical volume $V_{c}$ introduced in
Eq. (\ref{Susc5}). $V_{c}$ is defined as the volume at which the relaxation
time (or the escape rate) is equal to the measuring time $\tau _{m}=100s$
(or measuring frequency). That is, if one defines the function 
\begin{equation}
F(V)=\lambda (\alpha ,\theta _{a},\theta _{m},\theta _{b})-\frac{\tau _{N}}{%
\tau _{m}},  \label{3}
\end{equation}
where $\theta _{a},\theta _{b},\theta _{m}$ correspond to the two minima and
maximum of the potential, respectively, the critical volume $V_{c}$ is
obtained as the volume that nullifies the function $F(V)$ for given values
of $T,H$ and all other fixed parameters ($\gamma ,\eta _{r},M_{s}$ and the
volume-distribution width $\sigma $).

Then, $M_{zfc}$ is defined according to Gittleman's model \cite{Gittleman},
namely 
\begin{equation}
Z\times M_{zfc}(H,T,\psi )=%
\displaystyle \int %
\limits_{0}^{V_{c}}{\cal D}V\,M_{sp}(H,T,V,\psi )+%
\displaystyle \int %
\limits_{V_{c}}^{\infty }{\cal D}V\,M_{b}(H,T,V,\psi )  \label{4app}
\end{equation}
where $M_{sp}$ and $M_{b}$ are the contributions to magnetization from
superparamagnetic particles with volume $V\leq V_{c}$ and particles still in
the blocked state with volume $V>V_{c}$. $f(V)=(1/\sigma \sqrt{2\pi })\exp
(-\log ^{2}(V/V_{m})/2\sigma ^{2})$, is the
log-normal volume distribution, $V_{m}$ being the mean volume; $Z\equiv \int_{0}^{\infty }{\cal D}%
V=\int_{0}^{\infty }Vf(V)dV.$

Eq.(\ref{4app}) can be rewritten as 
\begin{equation}
Z\times M_{zfc}=%
\displaystyle \int %
\limits_{0}^{V_{c}}{\cal D}V\,M_{sp}-%
\displaystyle \int %
\limits_{0}^{V_{c}}{\cal D}V\,M_{b}+%
\displaystyle \int %
\limits_{0}^{\infty }{\cal D}V\,M_{b}.  \label{5}
\end{equation}
Now using, 
\begin{equation}
M_{b}(H,T,V,\psi )=\frac{M_{s}^{2}H}{2K}\sin ^{2}\psi ,  \label{6}
\end{equation}
$M_{b}$ can be taken outside the integral in the last term above. Thus%
\footnote{%
The reason for doing so is to avoid computing the integral $%
\int_{V_{c}}^{\infty }$ which is numerically inconvenient.}, 
\[
Z\times M_{zfc}=%
\displaystyle \int %
\limits_{0}^{V_{c}}{\cal D}V\,(M_{sp}-\,M_{b})+\,Z\times M_{b}. 
\]

The final expression of $M_{zfc}$ is obtained by averaging over the angle $%
\psi $ ($\left\langle \sin ^{2}\psi \right\rangle =\frac{2}{3}$), 
\begin{equation}
M_{zfc}=\frac{1}{Z}%
\displaystyle \int %
\limits_{0}^{\pi /2}d\psi \sin \psi \times 
\displaystyle \int %
\limits_{0}^{V_{c}}{\cal D}V\,(M_{sp}-\,M_{b})+\,\frac{M_{s}^{2}H}{3K}.
\label{8}
\end{equation}
The expression of $M_{sp}$ varies according to the model used. Chantrell et
al. \cite{Chantrell et al.} have given an expression which is valid for $%
M_{s}HV/k_{B}T\ll 1,$%
\begin{equation}
M_{sp}(H,T,V,\psi )=\frac{M_{s}^{2}VH}{k_{B}T}\left( \cos ^{2}\psi +\frac{1}{%
2}\left[ 1-\cos ^{2}\psi (1-\frac{I_{2}}{I_{0}})\right] \right) ,  \label{9}
\end{equation}
with 
\begin{equation}
\frac{I_{2}}{I_{0}}=\frac{1}{\alpha }\left( -\frac{1}{2}+\frac{e^{\alpha }}{%
I(\alpha )}\right) ,\quad I(\alpha )=2%
\displaystyle \int %
\limits_{0}^{1}dxe^{\alpha x^{2}}.  \label{10}
\end{equation}
Note that upon averaging over $\psi ,$ the expression in (\ref{9}) reduces
to 
\begin{equation}
M_{sp}(H,T,V)=\frac{M_{s}^{2}VH}{3k_{B}T},  \label{11}
\end{equation}
which is just the limit of the Langevin function for $M_{s}HV\ll k_{B}T,$
i.e. 
\begin{equation}
M_{sp}(H,T,V)=M_{s}{\cal L}\left( \frac{M_{s}HV}{k_{B}T}\right) .
\label{11b}
\end{equation}

Therefore, the expression in (\ref{8}) becomes 
\begin{equation}
M_{zfc}=\frac{1}{Z}%
\displaystyle \int %
\limits_{0}^{V_{c}}{\cal D}V\,\left( M_{sp}-\,M_{b}\right) +\,\frac{%
M_{s}^{2}H}{3K}  \label{12app}
\end{equation}
This is valid only in the case of a relaxation time independent of $\psi ,$
as in the N\'{e}el-Brown model, which is applicable to an assembly of
uniformly oriented particles. However, if one wanted to use the expressions
of the relaxation time given by Coffey et al. and others, where $\tau $
depends on the angle $\psi ,$ as is the case in reality, one should not
interchange integrations over $\psi $ and $V,$ as is done in (\ref{8}),
since $V_{c}$ in general depends on $\tau $ and thereby on $\psi $.

Therefore, the final expression for $M_{zfc}$ that was used in our
calculations for determining the temperature $T_{\max }$ is given by eqs. (%
\ref{11}), (\ref{11b}), (\ref{12app}).

\end{document}